\documentstyle[prl,aps]{revtex} \def\narrowtext{} \tighten\twocolumn

\begin{document}
\draft
\title{
\begin{minipage}[t]{7.0in}
\scriptsize
\begin{quote}
\leftline{{\it Phys. Rev. Lett.}, in press (2000)}
\raggedleft{\rm NORDITA -- 1999/49 CM}\\
\raggedleft {\rm cond-mat/9908202}
\end{quote}
\end{minipage}
\medskip
\\Spin Wave Theory of Double Exchange Ferromagnets}
\author{D. I. Golosov\thanks{E-mail: golosov@nordita.dk}}
\address{NORDITA, Blegdamsvej 17, DK-2100 Copenhagen \O, Denmark}
\address{%
\begin{minipage}[t]{6.0in}
\begin{abstract}
We construct the $1/S$ spin-wave expansion for double exchange
ferromagnets at T=0. It is assumed that the value of Hund's rule
coupling, $J_H$, is sufficiently large, resulting in a fully
saturated, ferromagnetic half-metallic ground state.
We evaluate corrections to the magnon dispersion law, and we also find
that, in contrast to earlier statements in the literature,
magnon-electron scattering does give rise to spin wave damping.
We analyse the 
momentum dependence of these quantities and discuss the experimental
implications for colossal magnetoresistance compounds.  
\typeout{polish abstract}
\end{abstract}
\pacs{PACS numbers: 75.30.Vn, 75.30.Ds, 75.10.Lp}
\end{minipage}}

\maketitle
\narrowtext

  The \nolinebreak phenomenon of 
colossal magnetoresistance \linebreak(CMR), with its potential
technological applications, has motivated an extensive experimental and
theoretical research effort directed at the understanding of
properties of doped 
manganese oxides\cite{review}. In particular, considerable attention
has been paid 
to magnetic properties of these compounds. An ubiquitous feature
shared by the numerous theoretical models of the CMR compounds is
the presence of a strong ferromagnetic Hund's rule exchange coupling, $J_H$,
between the spins of itinerant $e_g$ electrons and the core spins of
${\rm Mn}$ ions. The 
kinetic energy of itinerant electrons is then minimized when the ionic
spins are parallel to each other; this
gives rise to the conduction-electron mediated {\it double exchange} 
ferromagnetism \cite{DeGennes}. The physics of double exchange is thus
completely different from that of both Heisenberg exchange and RKKY
interaction (the latter corresponding to the case of small $J_H$).

Since the value of core spin is relatively large, $S=3/2$, studying the
effects of double exchange interaction in the large-$S$
limit should provide at least a qualitative
description of the low-temperature properties of the manganites. 
At the same time, the quantum nature of the core spins does affect the
behaviour of the system in a profound way, as indicated by the
variational  results. At finite $S$, the continuum of
Stoner-like single-particle excitations with finite energies is
present even at $J_H \rightarrow \infty$ \cite{Okabe}; also, the spin wave
spectra obtained variationally show appreciable
quantum-spin corrections\cite{Okabe,Wurth}.
These conclusions are fully supported by the exact diagonalization
studies of one-dimensional double exchange 
systems\cite{Trugman,Kaplan,Kaplan2}. 

It is therefore somewhat surprising that, apart from the leading-order
calculations\cite{Nagaev70,Kubohata,Furukawa96,Nagaev98} (which do not
account for any 
quantum corrections), the perturbative spin wave theory of double
exchange magnets remains underdeveloped. The objective of the present
paper is to partially fill this gap. Our approach (unlike that of
Ref. \cite{Irkhin}) remains valid in the experimentally relevant case
of large carrier densities. We will show that the subleading 
terms in the 1/S expansion, which originate from  magnon-electron
scattering, provide corrections to the magnon dispersion law, and also
give rise to magnon damping. We will also see how the momentum
dependence of these quantities is affected by the Fermi surface
geometry. The experimental data will be discussed briefly. 

We start with the usual double exchange Hamiltonian,
\begin{equation}
{\cal H}=-\frac{t}{2} \sum_{\langle i,j \rangle,\alpha} 
( c^\dagger_{i \alpha}c_{j \alpha} +c^\dagger_{j \alpha}c_{i
\alpha} ) -
 \frac{J_H}{2S} \sum_{i, \alpha, \beta} \vec{S}_i
\vec{\sigma}^{\alpha \beta} c^\dagger_{i\alpha} c_{i\beta} \,.
\label{eq:Ham0}
\end{equation}
Here  $c_{j \alpha}$ are the electron annihilation operators,
$\vec{S}_i$ are the operators of the core (localized) spins
located at the sites of a square (or simple cubic) lattice, and
the vector $\vec{\sigma}^{\alpha \beta}$ is composed of Pauli
matrices. We assume, for simplicity, that  only one
conduction-electron orbital (with two possible values of spin
projection, $\alpha=\uparrow,\downarrow$) is available per each site
$i$. Throughout the paper we use units in which hopping $t$, $\hbar$,
and the lattice spacing are all equal to unity, and we consider the
$T=0$ case.

There is presently no reason to doubt that for $S \geq 3/2$  the
ground state of the Hamiltonian (\ref{eq:Ham0}) is ferromagnetic,
at least for an infinite system in two or three dimensions and with 
a finite number of electrons per site, $x<1$ (this is corroborated by
the variational calculations \cite{Okabe,Wurth,Edwards}; see also Ref.
\cite{Kubo}  for the 1D case, and Ref. \cite{Trugman} for finite-size
systems). To leading order in $1/S$, the electron spectrum in the
ferromagnetic state is given by $\epsilon_{\vec{k}}^{\uparrow,\downarrow}=\epsilon_{\vec{k}} \mp
J_H/2$, where for the tight-binding model of Eq. (\ref{eq:Ham0}),
$ \epsilon_{\vec{k}}=-\sum_{a=1}^d \cos k_a$, and 
$d$ is the dimensionality of the lattice\cite{spec}. We will only consider
the half-metallic case \cite{halfmetal} when the chemical potential
lies below the bottom of the
upper band, $\epsilon_F < \frac{1}{2}J_H - d+ {\cal O}(1/S)$, so that
only spin-up electrons are present in the ground state.

The magnon operators $a_i$ are introduced by means of the Holstein -- Primakoff
transformation (including the subleading terms) of the operators $\vec{S}_i$, 
and the canonical
transformation\cite{Nagaev98} ${\cal H} \rightarrow {\cal H}^\prime = 
\exp(-U) {\cal H} \exp(U)$ with
\begin{equation}
{\cal U}=\frac{J_H}{\sqrt{2SN}} \sum_{\vec{k},\vec{p}} \left(\frac 
{c^\dagger_{\vec{k} \uparrow} c_{\vec{k}+\vec{p} \downarrow}
a^\dagger_{\vec{p}}}{\epsilon^\uparrow_{\vec{k}} -
\epsilon^\downarrow_{\vec{k} + \vec{p}}} - {\rm h. c.} \right)\,
\label{eq:canonical}
\end{equation}
(where $N$ is the number of lattice sites and the summation is over
the first Brillouin zone) is then applied to the Hamiltonian, thereby
accomplishing the transition to the ``true magnons''. The resultant
Hamiltonian can be written as an expansion in powers of $S^{-1/2}$:
\begin{equation}
{\cal H}^\prime=\sum_{\vec{k},\alpha} \epsilon_{\vec{k}}^\alpha
c^\dagger_{\vec{k} 
\alpha} c_{\vec{k} \alpha} + {\cal H}^\prime_2+{\cal
H}^\prime_3+{\cal H}^\prime_4+...\,\,\,. 
\label{eq:Ham1}
\end{equation} 
Here, the term ${\cal H}^\prime_1 \propto S^{-1/2}$ has been
eliminated by the canonical transformation, and 
\begin{eqnarray}
&&{\cal H}_2^\prime=\frac{J_H}{4NS}\sum_{1,2,3,4}^{}{\!\!}^{'}\left\{
\frac{c^\dagger_{1\downarrow}c^\dagger_{2\uparrow}c_{3\uparrow}c_{4\downarrow}}{\epsilon_2^\uparrow-\epsilon_4^\downarrow}
+\frac{\epsilon_1- \epsilon_{1+3}}{\epsilon_{1}^\uparrow-
\epsilon_{1+3}^\downarrow}c^\dagger_{1 \uparrow} c_{2 \uparrow} 
a^\dagger_3
a_4\right.
\nonumber\\
&& \left.-\frac{\epsilon_{2-3}-\epsilon_2}{\epsilon_{2-3}^\uparrow
-\epsilon_2^\downarrow}
c^\dagger_{1 \downarrow} c_{2 \downarrow} a^\dagger_3 a_4 +
{\rm h.c.} \right\}-\frac{J^2_H}{2NS}\sum_{\vec{k},\vec{q}}
\frac{c^\dagger_{\vec{k}\downarrow} c _{\vec{k} \downarrow}}
{\epsilon_{\vec{q}}^\uparrow -\epsilon_{\vec{k}}^\downarrow} .
\label{eq:Ham2}
\end{eqnarray}
Here, the subscripts $1,2,..$ stand for $\vec{p}_1, \vec{p}_2,...$,
and $\sum^\prime$ means that the quasimomentum conservation law is enforced. 
The spin wave energy $\omega_{\vec{p}}$ is equal
to the on-shell value of the real part of the magnon self energy
(cf. Ref. \cite{Furukawa96}). The leading-order 
term in self energy, which originates from ${\cal H}^\prime_2$,
is real and coincides with the earlier results
\cite{Nagaev70,Furukawa96,Nagaev98}, which in the $J_H \rightarrow \infty$
limit reduce to 
\begin{equation}
\omega_{\vec{p}}^{(0)}=-(2S)^{-1}\int
n_{\vec{q}}(\epsilon_{\vec{q}}- \epsilon_{\vec{p}+\vec{q}})
d^dq/(2\pi)^d \,,
\label{eq:spectrum0}
\end{equation}
where $n_{\vec{q}}$ is the Fermi distribution function for the spin-up
electrons. Using the explicit form of $\epsilon_{\vec{q}}$, one
obtains a Heisenberg-like expression,
$\omega_{\vec{p}}^{(0)}=|E|(d+\epsilon_{\vec{p}})/2dS$.
Here, $E=\int \epsilon_{\vec{q}}n_{\vec{q}} d^dq/(2\pi)^d$ is the
total energy of electrons calculated with 
respect to the center of the lower band.

Evaluation of the higher-order terms in Eq. (\ref{eq:Ham1}) involves
repeated commutations of the operator ${\cal U}$ (see
Eq. (\ref{eq:canonical})) with the original Hamiltonian,
Eq. (\ref{eq:Ham0}). 
In order to calculate the leading correction to the energy of a single
magnon, we will need to collect
only the terms which contain neither $c_{\vec{k}\downarrow}$ nor
$c_{\vec{k} \downarrow}^\dagger$ and are quadratic in magnon operators.
Such terms do not occur in ${\cal
H}_3^\prime$, and we find
\begin{eqnarray}
&&{\cal H}_4^\prime \stackrel{\rm eff}{=}\frac{J_H^4}{32S^2N^2}\left\{
-\sum_{1\div
4}^{}{\!}^{'}\frac{c^\dagger_{1\uparrow}c_{2\uparrow}a^\dagger_3
a_4}{\epsilon^\uparrow_2-\epsilon^\downarrow_{2+4}} \sum_{\vec{q}}
\frac{1}{\epsilon^\uparrow_{\vec{q}}-\epsilon^\downarrow_{2+4}}
\right[ \frac{4}{J_H}\nonumber\\ 
&&+\left.\frac{3\epsilon^\uparrow_{\vec{q}}+\epsilon^\uparrow_1-
4\epsilon^\downarrow_{2+4}}{(\epsilon^\uparrow_1-\epsilon^\downarrow_{2+4})
(\epsilon^\uparrow_{\vec{q}}-\epsilon^\downarrow_{2+4})}\right] 
+\sum_{1\div
6}^{}{\!}^{'}\frac{c^\dagger_{1\uparrow}c^\dagger_{2\uparrow}c_{3\uparrow}
c_{4\uparrow}a^\dagger_5a_6}{(\epsilon^\uparrow_4-\epsilon^\downarrow_{4+6})
(\epsilon_2^\uparrow-\epsilon^\downarrow_{4+6})} 
\nonumber\\
&&\left.\times\left[\frac{4}{J_H}+\frac{2\epsilon^\uparrow_3+\epsilon^\uparrow_1+\epsilon_2^\uparrow
-3\epsilon^\downarrow_{1+5}-\epsilon^\downarrow_{4+6}}
{(\epsilon^\uparrow_1-\epsilon^\downarrow_{1+5})
(\epsilon^\uparrow_3-\epsilon^\downarrow_{1+5})}\right]\right\}  
+{\rm h.c.}\,.
\label{eq:Ham4}
\end{eqnarray}
Naturally, both ${\cal H}^\prime_2$ and ${\cal H}^\prime_4$ have 
well-defined $J_H \rightarrow \infty$ limits.
In what follows we will for
simplicity restrict ourselves to the case of infinite $J_H$;
straightforward generalization to the finite-$J_H$ case can always be
accomplished with the help of Eqs. (\ref{eq:Ham2}) and (\ref{eq:Ham4}).

The first quantum correction to the magnon self energy is proportional
to $S^{-2}$ and includes the first-order  perturbative
contribution from 
${\cal H}^\prime_4$ as well as second-order contribution
$\Sigma_2(0,\vec{p})$ from  ${\cal H}^\prime_2$ (corresponding to the
first diagram in Fig. \ref{fig:diagram} below)\cite{Nagaev69}:
\begin{eqnarray}
&&\omega^{(1)}_{\vec{p}}=\frac{1}{16 S^2} \int (3 \epsilon_{\vec{q}} - 4
\epsilon_{\vec{q}+\vec{p}})n_{\vec{q}}\frac{d^d q}{(2\pi)^d}+
{\rm Re}\Sigma_2(0,\vec{p})\,\,, \label{eq:real} \\
&&\Sigma_2(\omega,\vec{p})= \frac{1}{16S^2}\int\!
\frac{(\epsilon_{\vec{q}}+\epsilon_{\vec{k}}-2
\epsilon_{\vec{p}+\vec{q}})^2 (1-n_{\vec{k}})n_{\vec{q}}}
{\omega+\epsilon_{\vec{q}}-
\epsilon_{\vec{k}}+i0} \frac{d^dqd^dk}{(2\pi)^{2d}}.
\nonumber \\
&& \label{eq:sigma2}
\end{eqnarray}
For  the $x \ll 1$ case, this agrees with Ref. \cite{Irkhin}. 

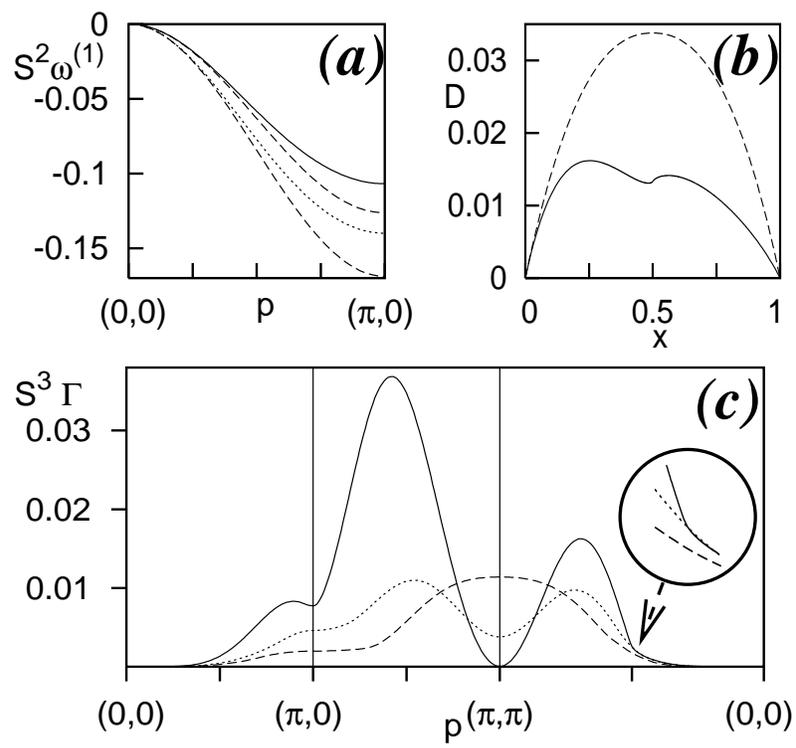
\begin{figure}
\caption{ {\it(a)} The leading $1/S$-corrections to the spin wave
energy in a 2D system with carrier concentration $x=0.3$ (solid line)
and $x=0.4$ (dotted 
line) as functions of momentum. The dashed lines
represent respective 
Heisenberg-like fits, $\tilde{\omega}^{(1)} S^2$. 
{\it (b)} Doping dependence of spin stiffness $D$ in a 2D system with
$S=3/2$. The dashed line corresponds to the classical value, $D^{(0)}$.  
{\it (c)} Momentum dependence of spin wave damping, $\Gamma(\vec{p})$,
for a 2D system with $x=0.49$, $x=0.4$, and $x=0.3$ (solid, dotted and
dashed lines, respectively). For $x=0.49$, note the $p=k_F$ anomaly,
visible at $\vec{p}\approx\{\pi/2, \pi/2\}$ (see the inset), and the
smallness of $\Gamma(2k_F)$.}  
\label{fig:spectrum}
\end{figure}

For a 2D system with band filling values $x=0.3$ and $x=0.4$,  the
quantity $\omega^{(1)}_pS^2$ is plotted in Fig. \ref{fig:spectrum}
{\it a}
(solid and dotted lines, respectively).
The dashed lines represent the corresponding  Heisenberg-like
fits,
 $\tilde{\omega}^{(1)}_{\vec{p}}S^2=2D^{(1)}(2+\epsilon_{\vec{p}})S^2$, 
where \cite{specific} 
\[
D^{(1)}=\frac{1}{4S^2}\int \frac{(\partial
\epsilon_{\vec{q}}/ 
\partial
q_x)^2}{\epsilon_{\vec{q}}-\epsilon_{\vec{k}}} 
(1-n_{\vec{k}})n_{\vec{q}}
\frac{d^dq d^dk}{(2\pi)^{2d}}-\frac{x|E|}{8dS^2}
\]
is the first quantum correction to spin stiffness.
The deviation of $\omega^{(1)}_{\vec{p}}$ from
$\tilde{\omega}^{(1)}_{\vec{p}}$ 
is not large, but it 
increases
dramatically 
for smaller $x$, in agreement with the 
numerical\cite{Kaplan} and variational\cite{Wurth} results.

The doping dependence of the spin stiffness, $D=D^{(0)}+D^{(1)}$, for
$S=3/2$ is shown in Fig. \ref{fig:spectrum} {\it b} (solid line). The
dashed line represents the leading-order term,
$D^{(0)}=|E|/8S$.  We see that for $S=3/2$ the value of $D^{(1)}$ 
is not small numerically\cite{stiff}, implying that quantum
corrections to the spin wave spectrum cannot be omitted in any
quantitative treatment\cite{Wurth,Kaplan}. Note that $D(x)$ is not symmetric
with respect to quarter filling, $x=0.5$, reflecting the loss of particle-hole
symmetry at finite $S$.

Spin wave damping is given by the on-shell value of the imaginary part
of magnon self energy, and the first-order perturbation theory
terms obviously do not contribute to it. It is also easy to see
that to second order in 
$1/S$, $\,{\rm Im} \Sigma_2(\omega_{\vec{p}},\vec{p})={\rm Im}
\Sigma_2(0,\vec{p})=0$. This is because the energies of all the intermediate
states occurring in the second-order perturbation theory terms are
higher than that of a single-magnon state (which to leading order
equals the ground state energy). Therefore, the integration contour in
the energy space terminates at the pole (cf. Eq. (\ref{eq:sigma2}) with
$\omega=0$), so
that the latter does not give rise to any imaginary contribution.
In the third order in $1/S$, a multitude of new diagrams
appear, combining all possible vertices from ${\cal
H}_2^\prime$, ${\cal H}_3^\prime$, and ${\cal H}_4^\prime$. By the
above reasoning, their leading-order imaginary parts vanish on
shell\cite{ladders},
with one exception. This exception is the second diagram
in Fig. \ref{fig:diagram}, which corresponds to the term with the {\it
second}-order pole in the standard third-order perturbation theory
formula\cite{Volume3}. This is nothing but a self energy correction
to $\Sigma_2(\omega_{\vec{p}},\vec{p})$ (Eq. (\ref{eq:sigma2})), and
the net third-order 
contribution to the magnon relaxation rate is thus given by the
diagram on the r.\ h.\ s. in Fig. \ref{fig:diagram}\cite{AFM}:
\begin{eqnarray}
\Gamma(\vec{p}) &=& \frac{\pi}{4 S^2} \int
 J(\vec{p},\vec{r})\frac{d^d r}{(2 \pi)^d}\,\,, \label{eq:imag} \\
J(\vec{p},\vec{r})& =& \int 
(\epsilon_{\vec{q}}-
\epsilon_{\vec{p}+\vec{q}})^2 (1-n_{\vec{q}+\vec{r}})n_{\vec{q}}
 \times \nonumber \\
&& \times \delta(\omega_{\vec{p}}^{(0)}-\omega_{\vec{p}-\vec{r}}^{(0)}+
\epsilon_{\vec{q}}-\epsilon_{\vec{q}+\vec{r}}) \frac{d^d q}{(2 \pi)^d}\,.
\label{eq:Jpq}
\end{eqnarray}
Note that $J(\vec{p},\vec{r})$
is of the order of $1/S$.

Thus in a double exchange system, the spin wave linewidth remains
finite even in a $J_H \rightarrow \infty$, $T=0$ case. While such a
possibility was discussed earlier\cite{Kaplan2} in connection with
recent experiments\cite{Lida}, it is at variance
with the prevalent view\cite{Furukawa97,Lida}. 
Magnon damping in a double exchange ferromagnet is due to magnon-electron 
scattering: a magnon can scatter into a lower-momentum state, while exciting
an electron from under the Fermi level (see Fig.
\ref{fig:diagram}). The momentum transfer into the
electronic subsystem allows us to speculate that  an  
electrical current may arise in this process. 

The value $\Gamma(\vec{p})$ for different values of bandfilling
$x$ in the 2D case is plotted in Fig. \ref{fig:spectrum} {\it c}. It shows
strong momentum dependence, reaching (for $S=3/2$) up to 10 \% of the
corresponding magnon energy. 
We note a difference from both a 
Heisenberg ferromagnet (with $\Gamma(\vec{p}) \equiv 0$ at $T=0$) 
and an RKKY ($s$-$f$ exchange) ferromagnet, where 
the leading term in $\Gamma(\vec{p})$ is threshold-like \cite{Vonsovskii}. 

We note that Eqs. (\ref{eq:imag}--\ref{eq:Jpq}) 
(as well as Eqs. (\ref{eq:canonical}--\ref{eq:sigma2})) remain 
unchanged if the first 
term in the Hamiltonian (\ref{eq:Ham0}) is written as
$\sum_{\vec{k},\alpha} \epsilon_{\vec{k}} c^\dagger_{\vec{k} \alpha}
c_{\vec{k} \alpha}$ with an arbitrary $\epsilon_{\vec{k}}$.
It is thus possible to use
Eqs. (\ref{eq:imag}--\ref{eq:Jpq}) to evaluate
$\Gamma(\vec{p})$ for an arbitrary single-band electron
dispersion. 
The overall profile of $\Gamma(\vec{p})$ (as well as that of
$\omega_{\vec{p}}$) is sensitive 
to both the carrier concentration value $x$ and 
the details of electron
bandstructure throughout the Brillouin zone. There are, however,
two generic features in the behaviour of $\Gamma(\vec{p})$:

\noindent {\bf 1. Long-wavelength limit.} When $p$ is small in
comparison to the Fermi momentum $k_F$,
$\Gamma(\vec{p})$ is proportional to
$p^5$ in the 2D case and to $p^6$ in three dimensions\cite{Chubukov}.
Here, the factor $p^2$ originates from the matrix element
$(\epsilon_{\vec{q}}-\epsilon_{\vec{p}+\vec{q}})^2$ in
Eq. (\ref{eq:Jpq}), and is multiplied by 
a $p^d$ corresponding to the space
volume available to the virtual magnon with momentum $\vec{p}-\vec{r}$
and energy $\omega_{\vec{p}-\vec{r}} < \omega_{\vec{p}} \approx
D^{(0)} p^2$ (see Fig. \ref{fig:diagram}).
The remaining factor of $p^1$ comes from electron kinematics.
In the one dimentional case, 
$\Gamma(p) \equiv 0$ everywhere at $p<k_F$.

\noindent {\bf 2. The anomaly at $p=k_F$}. The Fermi surface geometry
manifests itself through a weak singularity of $\Gamma(\vec{p})$ at
$p=k_F$\cite{location}. Indeed, the quantity 
$J(\vec{p},\vec{r})$ (see Eq. (\ref{eq:Jpq})) has a singularity at
$r=2k_F$, and vanishes at
$\omega^{(0)}_{\vec{p}-\vec{r}}=\omega^{(0)}_{\vec{p}}$. The anomaly in
$\Gamma(\vec{p})$ is due to the tangency
between these two surfaces (in $\vec{r}$-space), which occurs at
$p=k_F$.
At this point, the second (in 2D) or third (in 3D) derivative of
$\Gamma(p)$ suffers either a jump or a logarithmic divergence, depending
on the local geometry of these two tangent surfaces. These
singularities are too weak to be visible in Fig. \ref{fig:spectrum}
{\it c},
except for $x=.49$, when owing to the flatness of the Fermi surface
the singularity acquires a nearly-1D (jump in the {\it first}
derivative) character. 

The anomaly in $\Gamma(\vec{p})$ is of course accompanied by a
singularity in the third-order (in $1/S$) term in
$\omega_{\vec{p}}$. It is contained in the real part of the same
diagram, and for an isotropic dispersion law amounts
to a jump in the second derivative (2D) or to a logarithmic divergence in
either the first (1D) or third (3D) derivative of $\omega_{\vec{p}}$.

In view of the anticipated Fermi surface geometry of the CMR
compounds\cite{FSshape}, we mention the case when the Fermi
surface is nearly-cubic (nearly square). As the  
magnon isoenergetic surfaces are unlikely to have the same
shape\cite{curva}, this 
case is not reducible to 1D. In fact, when
$\vec{p}$ is perpendicular to a flat part of
the Fermi surface, the singular term in
$\Gamma(\vec{p})$ is proportional to $(p-k_F)^{3/2}\theta(p-k_F)$ in
2D and to $(p-k_F)^{2}\theta(p-k_F)$ in 3D \cite{location}. The 
singularity in $\omega_{\vec{p}}$ is given by
$-(p-k_F)^{3/2}\theta(p-k_F)$ in 2D and by $(p-k_F)^{2}{\rm
ln}|p-k_F|$ in 3D.

Some comments are in order concerning the experimental situation.
Low temperature spin dynamics, including spin wave damping and the
deviation of magnon spectrum from the Heisenberg-like form, is studied
intensively in both 3D perovskite\cite{Lida,Fernandez,Dai} and
quasi-2D layered \cite{Hirotastiff} CMR manganites. In both cases, 
unexpectedly large low-$T$ values of $\Gamma(p)$ at large $p$ were
found, which is consistent with our results (in some
compounds\cite{Dai}, $\Gamma(p)$ may be due in part to
electron-phonon coupling). 
However, the lack of detailed knowledge of electron
bandstructure restricts one's options in comparing theoretical and
experimental results quantitatively.

The low-momentum measurements of $\Gamma(\vec{p})$ in the perovskite
manganites\cite{Lida} yield the $p^4[T {\rm ln}( T/\omega_p)]^2$
dependence, attributable to the 
magnon-magnon scattering \cite{Kashcheev}.
We suggest that these measurements, both in 2D and 3D compounds, 
should be performed at lower temperatures in order to reveal the
$p^5$ or $p^6$ contribution of magnon-electron scattering (which is 
likely to have a numerically small prefactor\cite{small}). 
One should also try to identify the $p=k_F$ anomalies in $\Gamma$ and
$\omega$
(the irregularity seen in the data of Ref. \cite{Lida} at $p
\sim \pi/2$ may be a possible candidate).
We also suggest that a systematic study of doping dependence
of spin stiffness should be performed, in particular, in the layered
compounds\cite{HirotaTC}.

In conclusion, we have calculated the spin wave linewidth 
and correction to the spin wave energy  for 
a double exchange half-metallic ferromagnet.
The proposed new measurements would verify to what extent the
double exchange model accounts for the low-temperature spin dynamics of
the CMR compounds.

It is a pleasure to thank A. Luther for the many enlightening
discussions and steady encouragement. The author is grateful to A. V. Chubukov,
M. I. Kaganov,  T. A. Kaplan, and L.-J. Zou for their comments on the
present paper. 
Earlier discussions with
K. Levin, M. R. Norman, and R. Osborn are gratefully acknowledged.

{\it Note added in proof:} ~Our Eq. (\ref{eq:real}) agrees with
Ref. \cite{Auslender}, reporting a 1/S expansion for the
model with a strong ferromagnetic exchange, $J_f$, between the core
spins. However, the second-order calculation \cite{Auslender} in
itself is
not sufficient to evaluate the magnon damping at $J_f=0$, given by
Eq. (\ref{eq:imag}) above.

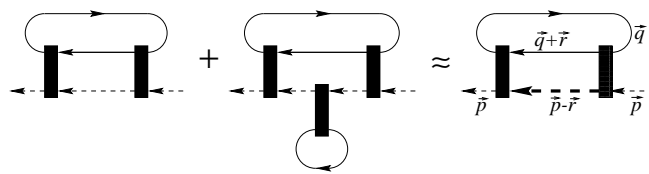
\begin{figure}
\caption{
Magnon self energy diagrams. Dashed and solid lines are the unperturbed
Green's functions of magnons, $(\omega+i0)^{-1}$, and of the spin-up
electrons, respectively. The bold dashed line is the exact magnon
Green's function, which to required accuracy is given by
$(\omega-\omega^{(0)}_{\vec{p}}+i0)^{-1}$. The boxes represent the
interaction, {\it i.e.} the $c^\dagger_\uparrow c_\uparrow a^\dagger
a$-terms in Eq. (\ref{eq:Ham2}).}
\label{fig:diagram}
\end{figure}


\begin{references}
\bibitem{stiff} A variational result\cite{Okabe} for $D^{(1)}$ in a 3D
system is smaller.
\bibitem{ladders} We note that the imaginary parts of the third-order
(in ${\cal H}_2^\prime$) ladder diagrams for the self energy behave as
$S^{-3}\omega
{\rm ln} \omega$, with the overall sign depending on momentum and on 
the bandstructure. As for ${\rm Im} \Sigma_2 (\omega,\vec{p})$, it is
proportional to $S^{-2}\omega \theta(\omega)$,
and thus has to be included in the third-order (in $1/S$) on-shell
calculation (see Fig. \ref{fig:diagram}).
\bibitem{Chubukov} After the present work was completed, we learned
that this result agrees with the calculations of A. V. Chubukov and
N. Shannon (unpublished, 1999).
\bibitem{location} Strictly speaking,
the anomaly is located at a given point $\vec{p}$ of the Fermi surface
only when the electron velocity  at this point is parallel to the magnon one, 
$\partial \epsilon_{\vec{q}}/\partial \vec{q} \parallel
\partial
\omega^{(0)}_{\vec{q}}/\partial \vec{q}$ at $\vec{q}=\vec{p}$.
This holds
along any symmetry direction, or for the tight-binding
model of Eq. (\ref{eq:Ham0}).
\bibitem{curva} This is because the bandstructure proposed in
Ref. \cite{FSshape} is not described by the tight-binding model of Eq. (\ref{eq:Ham0}).
\bibitem{small} For $\epsilon_{\vec{k}} \propto k^2$ and $p \ll k_F$,
we find $\Gamma = D k_F p^5/(90 \pi^3 S^2)$ in 2D and
(cf. Ref. \cite{Auslender}) $\Gamma= D k_F^2
p^6/(1680 \pi^3 S^2)$ in 3D.
\bibitem{HirotaTC} Our $D(x)$ in Fig. \ref{fig:spectrum} {\it b}
may be not inconsistent with 
$T_C(x)$ measured by M. Kubota {\it et al}  
(J. Phys. Chem. Solids {\bf 60}, 1161 (1999)), which has a maximum at
$x \approx .37$.
\end{references}
\end{document}